\documentclass[twoside]{article}

\usepackage[preprint]{aistats2026}
\usepackage[round]{natbib}

\usepackage{amsmath,amssymb,amsthm,mathtools,bm}
\usepackage{xspace}
\usepackage{booktabs}
\usepackage{graphicx}
\usepackage{pgf}

\usepackage{enumitem}
\usepackage[hypertexnames=false]{hyperref}
\usepackage{xcolor}
\usepackage{float}

\newtheorem{theorem}{Theorem}
\newtheorem{proposition}[theorem]{Proposition}

\theoremstyle{definition}

\newtheorem{assumption}[theorem]{Assumption}
\theoremstyle{remark}

\newcommand{\M}{\mathcal M}
\newcommand{\V}{\mathcal V}
\newcommand{\Pcal}{\mathcal P}
\newcommand{\E}{\mathbb E}
\newcommand{\R}{\mathbb R}
\newcommand{\KL}{\mathrm{KL}}

\newcommand{\ri}{\mathrm{ri}}
\newcommand{\conv}{\mathrm{conv}}

\newcommand{\tr}{\mathrm{tr}}

\newcommand{\norm}[1]{\left\lVert #1\right\rVert}

\newcommand{\GID}{\textsc{gid}\xspace}
\begin{document}

\runningtitle{Geometric Information Decomposition on the Sphere}

\runningauthor{You and Cho}

\twocolumn[

\aistatstitle{Geometric Information Decomposition for Weighted Empirical Measures on the Sphere}

\aistatsauthor{
	Kisung You
	\And
	Boram Cho
}

\aistatsaddress{
	Baruch College
	\And
	Yale University
}]
%\aistatsaddress{ Baruch College \And The Graduate Center, City University of New York } ]

\begin{abstract}
Weighted observations on the unit sphere arise in importance sampling, quadrature, and attention-weighted embeddings. Directional uncertainty is often summarized through a von Mises-Fisher (vMF) fit and its concentration or entropy. This summary uses only mean-direction information. It can miss antipodal, axial, girdle-like, or multimodal structure. We introduce geometric information decomposition (GID), which fits nested maximum-entropy projections to spherical features. Each gap measures the entropy reduction contributed by one feature level. The first gap is the fitted vMF distribution's KL divergence from uniformity. The second measures residual quadratic information, including Fisher-Bingham anisotropy. Later gaps describe finer angular structure. We establish invariance, consistency, alternative-regime asymptotic normality, and quadratic-form null calibration. Circular and spherical experiments include importance-weight calibration and a query-weighted digit projection. The results separate settings where vMF uncertainty is adequate from settings with higher-order structure.
\end{abstract}

% =============================================================
\section{INTRODUCTION}

Consider a probability measure on the $(d-1)$-dimensional sphere $S^{d-1}\subset\mathbb{R}^d$. Weighted support points represent this measure:
\begin{equation}
  \widehat P_w = \sum_{i=1}^n w_i \delta_{x_i},
  \quad x_i\in S^{d-1},~w_i\ge 0,~\sum_{i=1}^n w_i=1.
  \label{eq:weighted_empirical}
\end{equation}
The weights define integration under the represented law. Uniform weights $w_i=1/n$ give the usual empirical distribution. Nonuniform weights arise from posterior particles, importance samples, and quadrature rules. The same representation covers reliability, survey, and attention weights. For \eqref{eq:weighted_empirical}, our goal is to summarize the law's geometric variability on $S^{d-1}$.

A common model-based summary fits a von Mises-Fisher (vMF) distribution \citep{fisher_1953_DispersionSphere}. With
\[
  \widehat r=\sum_{i=1}^n w_i x_i,
  \qquad \widehat R=\norm{\widehat r},
\]
the fit maps $\widehat R$ to a concentration $\widehat\kappa$. It also gives fitted entropy or Kullback-Leibler (KL) divergence from uniformity. Large $\widehat\kappa$ indicates concentration around one direction. A value near zero means that the fitted vMF law is nearly uniform.

\begin{figure*}[t]
\centering
%\includegraphics[width=0.90\textwidth]{gid_figures/fig1_geometry_fingerprint.pdf}
%\resizebox{0.90\textwidth}{!}{\input{gid_figures/fig1_geometry_fingerprint.pgf}}
\input{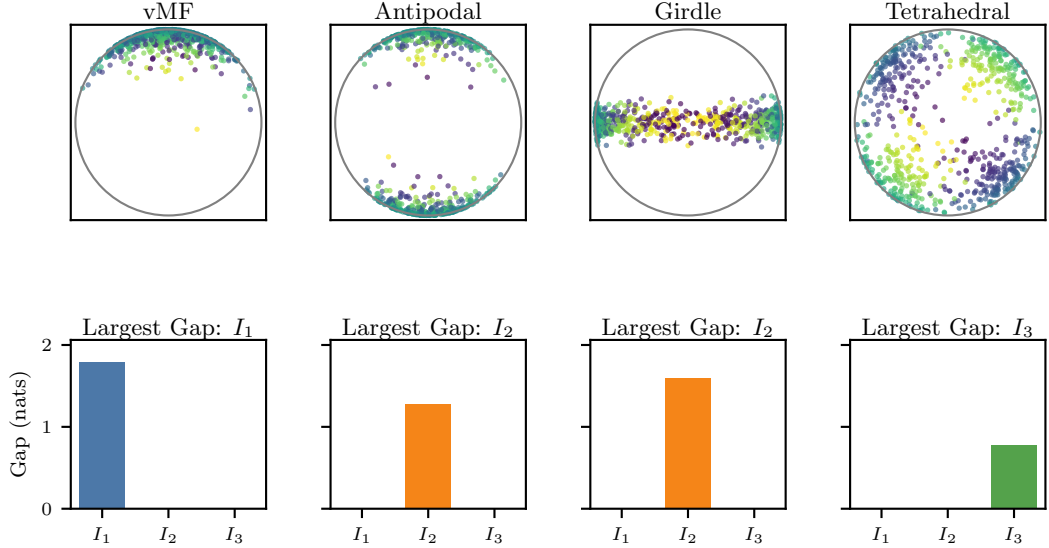}
\caption{Geometry-to-information fingerprint on $S^2$. The top row shows representative point clouds under orthographic projection. The bottom row reports the first three information gaps. vMF structure is first-order. The antipodal and girdle examples are second-order. The tetrahedral example first appears at a higher harmonic level.}
\label{fig:hero}
\end{figure*}

This first-order summary fails in several basic cases. Suppose $P$ places equal mass near $\mu$ and $-\mu$. Then $\E_P[X]\approx0$, so the fitted vMF model is nearly uniform. The distribution nevertheless remains concentrated around an axis. Equatorial concentration also yields a vanishing mean direction despite strong girdle structure. Symmetric modes can likewise cancel low-order moments while remaining far from uniform. In each case, $\widehat\kappa$ confuses a missing mean direction with high uncertainty. Figure~\ref{fig:hero} summarizes these cases and a tetrahedral example.

GID replaces the single first-moment fit with nested maximum-entropy projections. Let $\V_L$ denote the spherical feature space at level $L$. One choice is the span of spherical harmonics through degree $L$. Define $p_L^P$ as the maximum-entropy density whose $\V_L$ moments match those of $P$. Its entropy deficit from uniformity is
\[
  D_L(P)=\KL(p_L^P\sigma\,\|\,\sigma),
\]
where $\sigma$ is normalized surface measure. This deficit measures non-uniform information explained through level $L$. The increment
$
  I_L(P)=D_L(P)-D_{L-1}(P)
$
measures the information added by level $L$. We call the profile vector
\begin{equation*}
(I_1(P), I_2(P), \ldots,I_L(P))
\end{equation*}
the \emph{geometric information decomposition} (\GID). Level $1$ recovers vMF mean-direction information. Level $2$ captures residual quadratic anisotropy, including antipodal, elliptical, and girdle structure. Higher harmonic levels describe finer angular patterns, including multimodality. The full hierarchy targets low- and moderate-dimensional directional data. Complete degree-two and higher spaces can be impractical in large dimensions. Section~\ref{sec:computation} therefore considers structured and projected features.

The same construction extends to compact manifolds. Below, $\nu$ denotes normalized Riemannian volume, with $\nu=\sigma$ on the sphere. The identity
\[
  I_L(P)=\KL(p_L^P\nu\,\|\,p_{L-1}^P\nu)
\]
specializes the information-geometric Pythagorean identity to nested maximum-entropy projections. We use it as an estimable uncertainty decomposition for measures represented by \eqref{eq:weighted_empirical}.

\paragraph{Contributions.}
GID treats the vMF entropy deficit as the first gap rather than a complete uncertainty summary for measures in \eqref{eq:weighted_empirical}. Nested maximum-entropy projections yield a hierarchical KL decomposition for spherical features. We establish basis and rotation invariance, consistency, alternative-regime normal limits, and second-order null calibration. Low-order interpretations and dimensional limits motivate structured or projected features for embeddings. Experiments on $S^1$ and $S^2$ identify structure that vMF uncertainty misses. Illustrative code is available at \url{https://github.com/kisungyou/GeoInfoDec}.

% =============================================================
\section{RELATED WORK}

\paragraph{Maximum entropy and information geometry.}
The maximum-entropy principle selects the least informative distribution under specified constraints \citep{jaynes_1957_InformationTheoryStatistical}. Moment-constrained maximum entropy is dual to exponential-family likelihood fitting. Its KL geometry is classical. \citet{csiszar_1975_$I$DivergenceGeometryProbability} developed information projection. \citet{amari_2007_MethodsInformationGeometry} and \citet{amari_2001_InformationGeometryHierarchy} studied Pythagorean decompositions for nested exponential families. Our KL-gap identity applies this framework to weighted empirical measures on the sphere. The resulting profile supports directional interpretation, estimation, and null calibration. Standard exponential-family theory supplies the boundary and existence results \citep{brown_1986_FundamentalsStatisticalExponential,csiszar_2003_InformationProjectionsRevisited}.

\paragraph{Directional distributions.}
Directional statistics provides the low-order models used in our hierarchy \citep{mardia_2000_DirectionalStatistics}. The vMF distribution is the canonical first-order spherical model. It is widely used for normalized observations and spherical clustering \citep{fisher_1953_DispersionSphere,banerjee_2005_ClusteringUnitHypersphere}. Bingham, Fisher-Bingham, Kent, and related families model second-order directional structure. \citet{beran_1979_ExponentialModelsDirectional} studied exponential models containing the Fisher/vMF and Bingham cases. \citet{kent_1982_FisherBinghamDistributionSphere} introduced the constrained five-parameter Kent family on $S^2$. Our full linear-quadratic level is the eight-parameter Fisher-Bingham family. Within GID, these levels separate mean-direction, axial or girdle, and higher-order structure.

\paragraph{Higher-order models, computation, and diagnostics.}
Harmonic exponential families model angular structure on compact homogeneous manifolds \citep{cohen_2015_HarmonicExponentialFamilies}. Our high-order spherical construction is related but emphasizes interpretation of the gap profile. Second-order spherical families have nontrivial normalizing constants. Saddlepoint approximations are one computational option for Bingham-type models \citep{kume_2005_SaddlepointApproximationsBingham}. Goodness-of-fit and uniformity tests motivate our residual diagnostics \citep{jupp_2008_DatadrivenSobolevTests,garcia-portugues_2018_OverviewUniformityTests}. A low-order entropy summary is useful only when its fitted projection represents the observed measure adequately.

% =============================================================
\section{SETUP}

Let $M$ be a connected compact Riemannian manifold without boundary. Let $\nu$ denote normalized Riemannian volume, so $\nu(M)=1$. All densities are defined with respect to $\nu$. Compactness is not required by every definition. It provides a canonical uniform reference and avoids integrability complications. We write $\Pcal(M)$ for the Borel probability measures on $M$.

Let
\[
  \V_0 \subset \V_1 \subset \cdots \subset \V_L \subset L^2_0(M,\nu)
\]
be nested finite-dimensional spaces of continuous, mean-zero functions. Here, $L^2_0$ contains functions with zero $\nu$-mean. The zero level $\V_0=\{0\}$ corresponds to the uniform distribution. Let $q_L=\dim(\V_L)$, and let $\phi_L:M\to\R^{q_L}$ collect a basis for $\V_L$. For $P\in\Pcal(M)$, define
\[
  m_L(P)=\int_M \phi_L(x)\,dP(x).
\]
For the empirical measure in \eqref{eq:weighted_empirical}, the plug-in moment is
\[
  \widehat m_L=m_L(\widehat P_w)=\sum_{i=1}^n w_i\phi_L(x_i).
\]
This is the default empirical integration rule. Unless stated otherwise, all empirical moments, objectives, estimates, and calibrations use $\widehat P_w$. We therefore omit the qualifier \emph{weighted} below.

The feasible moment body is
\[
  \M_L=\conv\{\phi_L(x):x\in M\}\subset\R^{q_L}.
\]
A boundary moment can force a degenerate solution. The main results therefore assume that each relevant moment lies in $\ri(\M_L)$.

For $m\in\ri(\M_L)$, define the maximum-entropy density
\[
\begin{aligned}
  p_{L,m}
  =\arg\max_{p}\;& -\int_M p\log p\,d\nu \\
  \text{s.t. }& \int_M p\,d\nu=1,\quad
    \int_M \phi_Lp\,d\nu=m .
\end{aligned}
\]
When $m=m_L(P)$, write $p_L^P=p_{L,m_L(P)}$. When $m=\widehat m_L$, write $\widehat p_L=p_{L,\widehat m_L}$. These densities require their corresponding moments to lie in the stated relative interiors. The same restriction applies to all functionals below.

The dual log-partition function is
\[
  \psi_L(\lambda)=\log\int_M \exp\{\lambda^\top\phi_L(x)\}\,d\nu(x).
\]
The maximum-entropy solution has the exponential-family form
\[
  p_{L,m}(x)=\exp\{\lambda_L(m)^\top\phi_L(x)-\psi_L(\lambda_L(m))\},
\]
where $\lambda_L(m)$ solves
\[
  \nabla\psi_L(\lambda_L(m))=m.
\]
Equivalently, $\lambda_L(m)$ maximizes
\[
  \lambda^\top m-\psi_L(\lambda).
\]
Thus, the plug-in estimator $\widehat p_L$ maximizes the weighted empirical log score
\begin{align*}
  \widehat\lambda_L
  &= \underset{{\lambda\in\R^{q_L} }}{\arg\max}
  \left\{\lambda^\top\widehat m_L-\psi_L(\lambda)\right\} \\
  &=\underset{\lambda}{\arg\max}  \sum_{i=1}^n w_i\log p_{\lambda,L}(x_i).
\end{align*}
For i.i.d. observations with uniform weights, this objective is proportional to an ordinary log-likelihood. The same holds for weights proportional to integer frequency counts. Other weights define a weighted log-score or pseudo-likelihood M-estimator.

\section{GEOMETRIC INFORMATION DECOMPOSITION}

\subsection{Entropy deficits and effective uncertainty}

Because $\nu$ is normalized, the uniform density $p\equiv1$ has zero entropy. Every density has nonpositive entropy:
\[
  h(p)=-\int_M p\log p\,d\nu\le 0.
\]
We define the level-$L$ entropy deficit from uniformity by
\[
  D_L(P)=\KL(p_L^P\nu\,\|\,\nu)
  =\int_M p_L^P\log p_L^P\,d\nu.
\]
The corresponding effective uncertainty is
\[
  U_L(P)=\exp\{-D_L(P)\}\in(0,1].
\]
The value $U_L(P)=1$ means that the matched features carry no non-uniform information. Smaller values indicate more concentration or explained structure.

The level-$L$ information gap is
\[
  I_L(P)=D_L(P)-D_{L-1}(P),\qquad L\ge 1.
\]

The empirical versions are
\begin{align}
  \widehat D_L&=D_L(\widehat P_w),
  &\widehat U_L&=\exp\{-\widehat D_L\},\notag\\
  \widehat I_L&=\widehat D_L-\widehat D_{L-1}.
  \label{eq:empirical_versions}
\end{align}

\subsection{Main structural properties}

The next theorem states the Pythagorean identity for nested maximum-entropy models \citep{csiszar_1975_$I$DivergenceGeometryProbability, amari_2001_InformationGeometryHierarchy, amari_2007_MethodsInformationGeometry}. This identity motivates a spherical uncertainty diagnostic for \eqref{eq:weighted_empirical}.

\begin{theorem}[Monotonicity and KL-gap identity]
\label{thm:kl_gap}
Assume $m_L(P)\in\ri(\M_L)$ for $L=0,1,\ldots,L_{\max}$. Assume also that the feature spaces are nested. Then
\[
  0=D_0(P)\le D_1(P)\le\cdots\le D_{L_{\max}}(P).
\]
Moreover, for each $L\ge 1$,
\begin{equation}
  I_L(P)=D_L(P)-D_{L-1}(P)
  =\KL(p_L^P\nu\,\|\,p_{L-1}^P\nu).
  \label{eq:kl_gap}
\end{equation}
Consequently,
\begin{equation}
  D_L(P)=\sum_{\ell=1}^L I_\ell(P).
  \label{eq:sum_gaps}
\end{equation}
\end{theorem}

Identity \eqref{eq:kl_gap} equates each entropy increment with the information distance between adjacent maximum-entropy projections. This motivates the term level-$L$ information gap.

\begin{theorem}[Basis invariance]
\label{thm:basis_invariance}
Let $\widetilde\phi_L=T\phi_L$ for an invertible matrix $T$. Whenever the relevant projections exist, $p_L^P$, $D_L(P)$, $U_L(P)$, and $I_L(P)$ are unchanged.
\end{theorem}

Basis invariance makes the decomposition depend on feature spaces rather than their coordinates.

\begin{theorem}[Isometry invariance]
\label{thm:isometry_invariance}
Let $\Gamma$ be a group of isometries of $M$. Suppose each $\V_L$ is invariant under composition with $g\in\Gamma$. Thus, $f\in\V_L$ implies $f\circ g^{-1}\in\V_L$. For any $P$ with the relevant projections and any $g\in\Gamma$,
\begin{align*}
  D_L(g_\# P)&=D_L(P), \\
  U_L(g_\# P)&=U_L(P), \\
  I_L(g_\# P)&=I_L(P).
\end{align*}
\end{theorem}

On the sphere, cumulative harmonic spaces $\V_L=\bigoplus_{\ell=1}^L\mathcal H_\ell$ give rotation invariance. Hence, the decomposition describes distributional geometry rather than coordinates.

% =============================================================
\section{ESTIMATION THEORY}

The quantities $D_L(P)$ and $I_L(P)$ are population functionals. Their plug-in estimates are defined in \eqref{eq:empirical_versions}. For descriptive analysis, $\widehat P_w$ itself is the target. For inference, $\widehat P_w$ is a random representation of an underlying law $P$. Calibration and testing require the second interpretation. Deterministic quadrature supports consistency but does not define sampling $p$-values by itself. The moment assumptions cover ordinary empirical measures, deterministic nonuniform weights, importance samples, and quadrature approximations. Each application must supply the relevant law of large numbers or central limit theorem. Throughout this section, estimators are extended arbitrarily off their empirical-interior events.

\begin{assumption}[Interior and identifiability]
\label{assump:interior}
For the levels under consideration, $m_L(P)\in\ri(\M_L)$. After removing linear dependencies that are constant $\nu$-almost surely, the exponential family generated by $\phi_L$ is minimal.
\end{assumption}

\begin{assumption}[Moment convergence]
\label{assump:moment_conv}
For each fixed $L$,
\[
  \widehat m_L=\sum_{i=1}^n w_i\phi_L(x_i)\xrightarrow{p}m_L(P).
\]
\end{assumption}

\begin{theorem}[Consistency]
\label{thm:consistency}
Under Assumptions \ref{assump:interior} and \ref{assump:moment_conv}, for each fixed $L$,
\[
  \widehat\lambda_L\xrightarrow{p}\lambda_L(P),\qquad
  \widehat D_L\xrightarrow{p}D_L(P),\qquad
  \widehat I_L\xrightarrow{p}I_L(P).
\]
\end{theorem}

For asymptotic normality, suppose the plug-in moments satisfy
\begin{equation}
  a_n\{\widehat m_L-m_L(P)\}\Rightarrow N(0,\Sigma_L),
  \label{eq:moment_clt}
\end{equation}
for some $a_n\to\infty$. Consider i.i.d. observations $X_i\sim P$ with deterministic, observation-independent weights. Then $a_n=(\sum_i w_i^2)^{-1/2}$ under a Lindeberg condition such as
\[
  \frac{\max_i w_i}{(\sum_j w_j^2)^{1/2}}\to 0.
\]
The quantity $(\sum_iw_i^2)^{-1}$ is the Kish weight effective sample size. It gives this CLT scaling only in the stated setting. The theorem itself requires only \eqref{eq:moment_clt}.

Let
\[
  D_L(m)=\sup_{\lambda}\{\lambda^\top m-\psi_L(\lambda)\}.
\]
Then $D_L(P)=D_L(m_L(P))$ and
\[
  \nabla_m D_L(m)=\lambda_L(m).
\]

\begin{theorem}[Delta-method asymptotics away from zero gaps]
\label{thm:clt}
Assume \ref{assump:interior} and the moment CLT \eqref{eq:moment_clt}. The event $\{\widehat m_L\in\ri(\M_L)\}$ then has probability tending to one. On this event,
\[
  a_n(\widehat D_L-D_L(P))
  \Rightarrow
  N(0,\lambda_L(P)^\top\Sigma_L\lambda_L(P)).
\]
For $I_L=D_L-D_{L-1}$, let $\Pi_{L-1}$ be the full-row-rank matrix satisfying $\phi_{L-1}(x)=\Pi_{L-1}\phi_L(x)$. Define
\[
  \gamma_L=
  \lambda_L(P)-\Pi_{L-1}^\top\lambda_{L-1}(P).
\]
Then
\begin{equation}
  a_n(\widehat I_L-I_L(P))
  \Rightarrow
  N(0,\gamma_L^\top\Sigma_L\gamma_L).
  \label{eq:I_clt}
\end{equation}
\end{theorem}

Theorem~\ref{thm:clt} applies under the alternative. It does not calibrate a zero-gap test because the first derivative of $I_L$ vanishes under that null. Null calibration is therefore second order.

\begin{theorem}[Second-order null calibration]
\label{thm:null}
Write the level-$L$ feature vector as $\phi_L=(u,v)$. Here, $u$ spans $\V_{L-1}$ and $v$ spans the added coordinates. Thus, $q=q_L-q_{L-1}$. Suppose $I_L(P)=0$, equivalently $p_L^P=p_{L-1}^P$. Let $\theta_0=(\alpha_0,0)$ be the corresponding level-$L$ natural parameter. Define
\[
  \mathcal J_0=\mathrm{Var}_{\theta_0}\{(u(X),v(X))\}
  =\begin{pmatrix} J_{uu} & J_{uv}\\ J_{vu} & J_{vv}\end{pmatrix}.
\]
Under the minimality condition in Assumption~\ref{assump:interior}, $J_{uu}$ is positive definite. The Schur complement
\[
  \mathcal S=J_{vv}-J_{vu}J_{uu}^{-1}J_{uv}
\]
is also positive definite. If
\[
  a_n\{\widehat m_L-m_L(P)\}\Rightarrow Z\sim N(0,\Sigma_L),
\]
then
\begin{equation}
  a_n^2\widehat I_L
  \Rightarrow
  \frac12 Z^\top \mathcal{N}_0^\top \mathcal S^{-1}\mathcal{N}_0Z,
  \quad
  \mathcal{N}_0=\begin{pmatrix}-J_{vu}J_{uu}^{-1} & I_q\end{pmatrix}.
  \label{eq:null_quad}
\end{equation}
Equivalently, the limit is a weighted sum of independent $\chi^2_1$ variables. Its coefficients are the eigenvalues of $\frac12 \mathcal S^{-1/2}\mathcal{N}_0\Sigma_L\mathcal{N}_0^\top \mathcal S^{-1/2}$. Now suppose $X_i$ are i.i.d. from the reduced maximum-entropy model with uniform weights. Then $\Sigma_L=\mathcal J_0$, $a_n=\sqrt n$, and
\[
  2a_n^2\widehat I_L\Rightarrow \chi^2_q.
\]
Inequality or cone restrictions can instead produce a chi-bar-square limit \citep{self_1987_AsymptoticPropertiesMaximum}.
\end{theorem}

Theorem~\ref{thm:clt} gives standard errors for nonzero population gaps. Testing a zero gap at a new level requires Theorem~\ref{thm:null}, its sandwich form, or a bootstrap imposing the moment null.

\paragraph{Operational null calibration.}
Theorem~\ref{thm:null} gives a direct plug-in procedure. First, fit the reduced model and form $\widehat\theta_0=(\widehat\alpha,0)$. Estimate
$
  \widehat{\mathcal J}_0=\mathrm{Var}_{\widehat\theta_0}\{\phi_L(X)\}
$
by quadrature, Monte Carlo, or analytic moments. Partition this matrix into $\widehat J_{uu},\widehat J_{uv},\widehat J_{vu},\widehat J_{vv}$. Then compute $\widehat{\mathcal S}$ and $\widehat{\mathcal N}_0$. Estimate $\Sigma_L$ under the actual sampling mechanism. For i.i.d. samples with deterministic weights and $a_n=(\sum_iw_i^2)^{-1/2}$, use
\[
  \widehat\Sigma_L=\frac{\sum_iw_i^2\{\phi_L(x_i)-\widehat m_L\}\{\phi_L(x_i)-\widehat m_L\}^\top}{\sum_iw_i^2}.
\]
For self-normalized importance weights with $a_n=\sqrt n$, use the influence estimate
\[
  \widehat\Sigma_L=\frac1n\sum_{i=1}^n (nw_i)^2
  \{\phi_L(x_i)-\widehat\mu_{0,L}\}\{\phi_L(x_i)-\widehat\mu_{0,L}\}^\top,
\]
where $\widehat\mu_{0,L}=\E_{\widehat\theta_0}\phi_L(X)$. This centering applies only under the null. Away from the null, use the alternative-regime influence function or a design-respecting bootstrap. A fitted-model parametric bootstrap requires correct specification. The null reference law is
\begin{equation}
  \sum_{j=1}^q \widehat\omega_j\chi^2_{1j},
  \quad
  \widehat\omega_j=\mathrm{eig}_j\left(\frac12\widehat{\mathcal S}^{-1/2}\widehat{\mathcal N}_0\widehat\Sigma_L\widehat{\mathcal N}_0^\top\widehat{\mathcal S}^{-1/2}\right),
  \label{eq:weighted_chisq_sim}
\end{equation}
Evaluate its upper tail at $a_n^2\widehat I_L$ numerically. With $B$ simulated draws $T_b^*$, use $\widehat p_B=\{1+\sum_b\mathbf1(T_b^*\ge a_n^2\widehat I_L)\}/(B+1)$. Let $B\to\infty$ when claiming a continuous-uniform limit. Valid calibration also requires $\widehat{\mathcal J}_0\xrightarrow{p}\mathcal J_0$, $\widehat\Sigma_L\xrightarrow{p}\Sigma_L$, and vanishing integration error. Under correct specification and uniform weights, this becomes the usual $\chi_q^2$ calibration for $2n\widehat I_L$.

% =============================================================
\section{THE SPHERE: FROM VMF TO HARMONIC STRUCTURE}\label{sec:sphere}

We return to $M=S^{d-1}$ with normalized surface measure $\sigma$. This section identifies the structure captured by fitted concentration and entropy. A spherical distribution may favor a direction, an axis, a great subsphere, or several separated modes. Each pattern has different implications for uncertainty. The harmonic hierarchy separates these patterns through progressively richer angular moments.

\subsection{Level 1: vMF as first-order uncertainty}

Let $\V_1$ be the span of linear coordinate functions $x\mapsto a^\top x$.  Then
\[
  p_1(x)=\exp\{\eta^\top x-\psi_1(\eta)\}.
\]
This is the vMF distribution with natural parameter $\eta=\kappa\mu$, where $\norm{\mu}=1$. The empirical mean resultant vector is
\[
  \widehat r=\sum_{i=1}^n w_ix_i,
  \qquad \widehat R=\norm{\widehat r}.
\]
When $0<\widehat R<1$, the fitted direction is $\widehat\mu=\widehat r/\widehat R$. The concentration $\widehat\kappa$ solves $A_d(\widehat\kappa)=\widehat R$. At $\widehat R=0$, $\widehat\kappa=0$ and the direction is unidentified. When $\widehat R=1$, no finite fit exists. The first entropy gap
\[
  \widehat I_1=\widehat D_1
\]
measures information explained by a nonzero mean direction. Thus, $\widehat\kappa$ summarizes only the hierarchy's first level.

When a vMF model fits well and residual checks support adequacy, $\widehat I_1$ dominates and vMF entropy is appropriate. A small $\widehat I_1$ indicates little first-order directional information. It does not imply proximity to uniformity.

The next proposition relates the first gap to mean resultant length near uniformity.
\begin{proposition}[Local form of the first-order gap]
\label{prop:I1_local}
Along any sequence of laws with $r(P)=\E_P[X]\to0$,
\[
  I_1(P)=D_1(P)=\frac d2\norm{r(P)}^2+o\{\norm{r(P)}^2\}.
\]
\end{proposition}

\subsection{Level 2: Fisher-Bingham-type structure}

Let $\V_2$ add traceless quadratic functions
\[
  x\mapsto x^\top Bx,
  \qquad B=B^\top,\quad \tr(B)=0.
\]
The level-2 density has the form
\[
  p_2(x)\propto \exp\{\eta^\top x+x^\top Bx\}.
\]
This family includes Fisher-Bingham and Bingham-type structure. The second gap
\[
  \widehat I_2=\widehat D_2-\widehat D_1
\]
measures the information added by second-order anisotropy.

This level first distinguishes directional uncertainty from axial uncertainty. In an equal antipodal mixture near $\mu$ and $-\mu$, the first moment cancels. Its matrix $\E[XX^\top]$ still has a dominant axis. For an equatorial girdle, the smallest-eigenvalue direction is normal to the girdle. Its orthogonal complement spans the girdle. A vMF fit misses both patterns, while $\widehat I_2$ detects them.

The usual second-moment matrix gives a direct interpretation. Let
\[
  Q(P)=\E_P[XX^\top],\qquad Q_0=I_d/d.
\]
In general, $I_2(P)=0$ exactly when $Q(P)=\E_{p_1^P}[XX^\top]$. If the first moment is zero, this condition reduces to $Q(P)=Q_0$. Nonzero traceless anisotropy then forces a positive gap. Near uniformity, the gap has the following local form.

\begin{proposition}[Local form of the second-order gap]
\label{prop:I2_local}
Along any sequence of laws with $\E_P[X]=0$ and $Q(P)\to Q_0$,
\[
  I_2(P)=\frac{d(d+2)}{4}\norm{Q(P)-Q_0}_F^2
  +o\{\norm{Q(P)-Q_0}_F^2\}.
\]
\end{proposition}
Thus, $I_2$ is locally equivalent to squared Frobenius anisotropy and retains a global entropy interpretation. Appendix~\ref{app:additional_figures} shows both local approximations.

\subsection{Higher levels: spherical harmonic exponential families}

Let $\{Y_{\ell r}:1\le r\le h_\ell\}$ be a real $L^2(\sigma)$-orthonormal basis for the degree-$\ell$ harmonic space $\mathcal H_\ell$. Set
$
  \V_L=\bigoplus_{\ell=1}^L \mathcal H_\ell.
$ Here $h_\ell=\binom{d+\ell-1}{\ell}-\binom{d+\ell-3}{\ell-2}$. Invalid lower indices are interpreted as zero. Then
\[
  p_L(x)=\exp\left\{\sum_{\ell=1}^L\sum_{r=1}^{h_\ell}\lambda_{\ell r}Y_{\ell r}(x)-\psi_L(\lambda)\right\}.
\]

Levels $L\ge3$ address structure that neither vMF nor Fisher-Bingham resolves. They can detect a small number of separated angular modes. Three symmetric modes on $S^1$ may have weak first-order and second-order signals. Four modes near tetrahedral directions on $S^2$ can behave similarly. Higher harmonic gaps reveal this residual non-uniformity. The hierarchy identifies which harmonic orders carry information when fitted concentration is small.

\paragraph{Adequacy and regularization.}
The gaps describe information explained by the chosen feature spaces. Residual diagnostics assess omitted structure. Invariant moment shrinkage handles boundary moments. Appendix~\ref{app:adequacy_regularization} presents MMD and harmonic diagnostics and separates exact GID from penalized fits.

% =============================================================
\section{COMPUTATION}\label{sec:computation}

The level-$L$ fit solves the concave dual objective $\ell_L(\lambda)=\lambda^\top\widehat m_L-\psi_L(\lambda)$. Its gradient is $\widehat m_L-\E_\lambda[\phi_L(X)]$. Its Hessian is $-\mathrm{Var}_\lambda\{\phi_L(X)\}$. Level 1 uses standard vMF solvers. Level 2 can use Fisher-Bingham or Bingham methods with saddlepoint normalizers \citep{kume_2005_SaddlepointApproximationsBingham}. Higher levels can use quadrature, spherical transforms, or Monte Carlo. Score matching can initialize parameters, but entropy requires $\psi_L$. Exact invariance assumes exact integration. A finite rule requires grid-size and rotation-sensitivity checks.

\paragraph{Scalability in ambient dimension.}
Full harmonic fitting is mainly suitable for low- and moderate-dimensional directional data. The complete degree-2 increment on $S^{d-1}$ adds
\[
  \dim\{\text{traceless quadratics}\}=d(d+1)/2-1
\]
features. Its normalizing constant is difficult to evaluate for large $d$. Thus, a complete $L=2$ fit is impractical for embeddings with hundreds of dimensions. High-dimensional variants should use structured feature spaces. Options include low-rank quadratic features after projection, diagonal or block-diagonal quadratics, random harmonic sketches, and application-specific invariant features. The same definitions and KL-gap theory apply when the selected structured subspaces are nested. GID then measures information only within those subspaces. Exact rotation invariance from Theorem~\ref{thm:isometry_invariance} requires the realized subspace to be invariant. A generic realized sketch is not invariant. Under a rotation-invariant sketching law, its gap may remain invariant in distribution. Suppose $\V_L^{\rm str}\subset\V_L^{\rm full}$ and both hierarchies share lower levels. Monotonicity gives $D_L^{\rm str}(P)\le D_L^{\rm full}(P)$. The difference measures information omitted by the structured features. Residual diagnostics such as \eqref{eq:harmonic_residual} can assess this loss. Repeated sketches or projections provide another check.

% =============================================================
\section{EXPERIMENTS}\label{sec:experiments}

The experiments test when vMF uncertainty is adequate or misleading. Comparisons include $\widehat R$, $\widehat\kappa$, and Frobenius anisotropy. Appendix~\ref{app:experiment_specs} gives the generating laws and numerical details. All gaps are in nats. The dominant gap is the largest raw estimate. This comparison is descriptive because null scale depends on increment dimension. Formal comparison requires Theorem~\ref{thm:null}.

\subsection{Controlled experiments on the circle}

On the circle, the hierarchy is a Fourier maximum-entropy model with features
\[
  \phi_L(\theta)=\{\cos(k\theta),\sin(k\theta):1\le k\le L\}.
\]
The Fourier hierarchy is fitted at $L=1,2,3,4$ on a 4096-point quadrature grid. Each row in Table~\ref{tab:s1} averages 100 replicates, each with $n=400$ observations. Parentheses contain Monte Carlo standard errors. The Pareto weights are independent of angle, have tail index $1.5$, and substantially reduce weight ESS.

\begin{table}[ht]
\centering
\caption{Fourier information gaps on $S^1$.}
\label{tab:s1}
\scriptsize
\resizebox{\columnwidth}{!}{%
\begin{tabular}{@{}llrrrrrr@{}}
\toprule
Scenario & Weights & Kish ESS & $I_1$ & $I_2$ & $I_3$ & $I_4$ & $D_4$\\
\midrule
vMF & uniform & 400 & 1.035 (0.004) & 0.002 (0.000) & 0.003 (0.000) & 0.003 (0.000) & 1.043 \\
vMF & Pareto & 90 & 1.060 (0.014) & 0.020 (0.004) & 0.016 (0.003) & 0.016 (0.003) & 1.111 \\
antipodal & uniform & 400 & 0.003 (0.000) & 0.726 (0.004) & 0.003 (0.000) & 0.009 (0.001) & 0.741 \\
antipodal & Pareto & 79 & 0.028 (0.005) & 0.750 (0.010) & 0.014 (0.002) & 0.022 (0.003) & 0.813 \\
trimodal & uniform & 400 & 0.003 (0.000) & 0.003 (0.000) & 0.534 (0.004) & 0.002 (0.000) & 0.542 \\
trimodal & Pareto & 92 & 0.026 (0.008) & 0.019 (0.004) & 0.556 (0.008) & 0.014 (0.002) & 0.615 \\
\bottomrule
\end{tabular}
}
\end{table}

The dominant gaps match the generating geometry. For a von Mises distribution, almost all information is first-order. In the antipodal mixture, $I_1$ is nearly zero while $I_2$ is large. Using vMF entropy as total uncertainty is therefore misleading. For the symmetric three-mode mixture, $I_3$ carries the signal. Unequal weights increase variability and upward bias in null gaps. The largest signal nevertheless remains stable.

Appendix~\ref{app:additional_figures} gives a separate trimodal illustration. The finite residual \eqref{eq:harmonic_residual} uses raw Fourier normalization, $a_\ell=1$, and $K=8$. It yields $R_{1,8}=0.421$, $R_{2,8}=0.420$, and $R_{3,8}=0.083$. The residual decreases when the third harmonic is included.

\subsection{Low-order experiments on the two-sphere}

The two-sphere experiment tests the first two levels. Level 1 uses linear features, while level 2 adds traceless quadratics. A deterministic 30,000-point Fibonacci rule approximates the log-partition function on $S^2$. Table~\ref{tab:s2} reports means and Monte Carlo standard errors from 30 uniform-weight replicates with $n=500$.

\begin{table}[ht]
\centering
\caption{Low-order information gaps and classical summaries on $S^2$. Here, $A_Q=\norm{\sum_iw_ix_ix_i^\top-I_3/3}_F$.}
\label{tab:s2}
\scriptsize
\resizebox{\columnwidth}{!}{%
\begin{tabular}{@{}lrrrrrr@{}}
\toprule
Scenario & $I_1$ & $I_2$ & $D_2$ & $\widehat R$ & $\widehat\kappa$ & $A_Q$\\
\midrule
vMF cloud & 1.780 (0.006) & 0.005 (0.001) & 1.786 & 0.876 & 8.07 & 0.550\\
antipodal mixture & 0.003 (0.001) & 1.490 (0.008) & 1.493 & 0.039 & 0.12 & 0.631\\
girdle distribution & 0.004 (0.001) & 0.919 (0.005) & 0.923 & 0.048 & 0.14 & 0.364\\
\bottomrule
\end{tabular}
}
\end{table}

These examples expose limitations in both vMF concentration and raw second-moment norms. The antipodal and girdle distributions have almost no mean direction. Their $\widehat R$ and $\widehat\kappa$ therefore suggest nearly uniform first-order uncertainty. The second gap detects their axial or equatorial structure. The comparison with $A_Q$ also shows why a raw second-moment norm cannot replace GID. A concentrated vMF cloud has large $A_Q$ because its first-order concentration induces anisotropic second moments. Its residual $I_2$ is nearly zero after this mean-direction information is fitted.

\subsection[Null calibration on S2]{Null calibration on $S^2$}

The calibration experiment uses a uniform target on $S^2$. It isolates the limiting calculation through the score surrogate $\widetilde I_2=\widehat m_v^\top\mathcal S^{-1}\widehat m_v/2$. This surrogate is second-order equivalent to $\widehat I_2$ under the null. Equal-weight samples compare $2n\widetilde I_2$ with $\chi^2_5$. Self-normalized samples use a vMF proposal with concentration $1.2$ and the plug-in quadratic form in \eqref{eq:weighted_chisq_sim}. Uniform $p$-values indicate calibration.

\begin{table}[ht]
\centering
\caption{Finite-sample calibration of the quadratic surrogate $\widetilde I_2$ on $S^2$ with $n=500$: 800 equal-weight and 350 importance-sampling replicates.}
\label{tab:calibration}
\scriptsize
\resizebox{\columnwidth}{!}{%
\begin{tabular}{@{}llrrrrrr@{}}
\toprule
Sampling & calibration & Kish ESS & $p_{.10}$ & $p_{.50}$ & $p_{.90}$ & size $.05$ & KS\\
\midrule
equal & $\chi^2_5$ & 500 & 0.121 & 0.521 & 0.890 & 0.043 & 0.037\\
importance & naive $\chi^2_5$ & 317 & 0.019 & 0.244 & 0.796 & 0.180 & 0.263\\
importance & quadratic form & 317 & 0.132 & 0.506 & 0.914 & 0.043 & 0.048\\
\bottomrule
\end{tabular}
}
\end{table}

First-order standard errors cannot calibrate null gaps. Informative weights require a sandwich covariance and a quadratic-form tail calculation. This experiment does not assess the finite-sample remainder $\widehat I_2-\widetilde I_2$.

\begin{figure}[th]
\centering
%\includegraphics[width=\columnwidth]{gid_figures/fig2_calibration_ecdf.pdf}
%\resizebox{0.90\textwidth}{!}{\input{gid_figures/fig1_geometry_fingerprint.pgf}}
\input{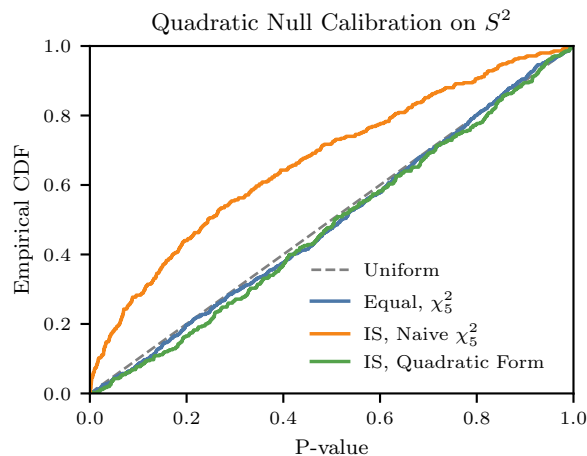}
\caption{Empirical CDFs of quadratic-surrogate calibration $p$-values for the three settings in Table~\ref{tab:calibration}.}
\label{fig:calibration_ecdf}
\end{figure}

Appendix~\ref{app:digits_table} reports a query-weighted digit illustration after projection to $S^2$.

% =============================================================
\section{DISCUSSION}

GID decomposes information captured by nested spherical maximum-entropy fits. Its gaps separate mean-direction information from residual quadratic and higher-order structure. Reports should include residual diagnostics, weight concentration, and sampling-design covariance. Zero-gap tests require second-order calibration, especially under informative weights. Alternative-regime standard errors do not apply at the null. Principled harmonic-order selection remains open.

% =============================================================

%\subsubsection*{Acknowledgements}
%All acknowledgments go at the end of the paper, including thanks to reviewers who gave useful comments, to colleagues who contributed to the ideas, and to funding agencies and corporate sponsors that provided financial support.  To preserve the anonymity, please include acknowledgments \emph{only} in the camera-ready papers. The acknowledgements do not count against the 9-page page limit in the camera-ready.

% If you use BibTeX in apalike style, activate the following line:
\bibliographystyle{apalike}
\bibliography{references}

%%%%%%%%%%%%%%%%%%%%%%%%%%%%%%%%%%%%%%%%%%%%%%%%%%%%%%%%%%%%

%\clearpage
\appendix
%\thispagestyle{empty}

% Supplementary material: To improve readability, you must use a single-column format for the supplementary material.
%\onecolumn
%\aistatstitle{Supplementary M}

\section{SUPPLEMENTARY EXPERIMENTS AND FIGURES}\label{app:supp_experiments}

\subsection{Audited experimental specifications}
\label{app:experiment_specs}

\paragraph{Circle experiments.}
The rerun for Table~\ref{tab:s1} uses seed 123, $n=400$, 100 replicates, and one 4096-point periodic grid at every level. The three laws are
\[
\begin{aligned}
 &\operatorname{vM}(0,4),\qquad
 \tfrac12\!\sum_{j=0}^1\operatorname{vM}(j\pi,8),\\
 &\tfrac13\!\sum_{j=0}^2\operatorname{vM}(2\pi j/3,12).
\end{aligned}
\]
For each sample, the uniform row uses $w_i=1/n$. The Pareto row draws mutually independent raw weights $W_i$ that are also independent of the angles. Their tails satisfy $\Pr(W_i>t)=t^{-1.5}$ for $t\ge1$, and their second moment is infinite. The normalized weights are $w_i=W_i/\sum_jW_j$. Thus, these rows are descriptive finite-sample stress tests rather than applications of the deterministic-weight CLT. A table entry is the replicate mean. Its parenthesized value is the replicate standard deviation divided by $\sqrt{100}$.

\paragraph{Two-sphere experiments.}
The rerun for Table~\ref{tab:s2} uses seed 123, $n=500$, 30 replicates, and one deterministic 30,000-point Fibonacci rule. Both levels use this rule. The vMF law has center $e_3$ and concentration 8. The antipodal law equally mixes vMF laws centered at $\pm e_3$, each with concentration 12. For the girdle, draw $\Theta\sim\operatorname{Unif}(0,2\pi)$ and $Z_0\sim N(0,0.195^2)$ independently. Set $Z=\min\{0.85,\max(-0.85,Z_0)\}$ and use
\[
 X=(\sqrt{1-Z^2}\cos\Theta,\sqrt{1-Z^2}\sin\Theta,Z).
\]
The fits are unpenalized and use BFGS with dual-gradient tolerance $10^{-6}$. A run is rejected when its final moment residual exceeds $10^{-5}$. No audited run was rejected, and no deficit or gap was clipped. Doubling the rule to 60,000 points changed every representative gap by at most $1.8\times10^{-7}$. Five random rotations of each of three representative samples changed every gap by at most $2.6\times10^{-6}$. The displayed precision is insensitive to these finite-rule effects. Exact rotation invariance remains a property of exact integration.

\paragraph{Calibration illustration.}
Table~\ref{tab:calibration} and Figure~\ref{fig:calibration_ecdf} use seed 123 and the quadratic surrogate $\widetilde I_2$. A 50,000-point Fibonacci rule evaluates $\mathcal S$. Each importance sample has 500 i.i.d. proposal draws from vMF$(e_3,1.2)$. Exact self-normalized ratios target the uniform law. Quadratic-form tails use 5,000 Monte Carlo draws, giving resolution $2\times10^{-4}$. Formal calculations should use an add-one exceedance correction. This correction changes no rounded table entry. The three rejection rates have binomial Monte Carlo standard errors of about 0.007, 0.021, and 0.011. The experiment checks covariance calibration of the null quadratic statistic. It does not assess the exactly fitted gap's finite-sample Taylor remainder.

\paragraph{Numerical inference.}
The theory assumes exact log-partitions and optimizers. Numerical deficit or gap error must be $o_p(a_n^{-1})$ for alternative-regime inference. For null inference based on the fitted gap, it must be $o_p(a_n^{-2})$. A fixed quadrature size guarantees neither rate. Negative gaps caused by numerical error must not be clipped for inference. Reports should include quadrature size, optimizer tolerance, moment residuals, and rotation sensitivity. Figure~\ref{fig:hero} used an archival ridge stabilization of $10^{-8}$. Removing it changes displayed gaps by less than $5\times10^{-9}$. No inferential result uses this stabilization.

\subsection{Additional diagnostics and figures}
\label{app:additional_figures}

These two figures support the main experiments but are not required for the central narrative. The specifications allow each example to be reproduced and interpreted independently.

\begin{figure}[ht]
\centering
%\includegraphics[width=0.95\columnwidth]{gid_figures/fig3_s1_hierarchy.pdf}
%\resizebox{0.90\textwidth}{!}{\input{gid_figures/fig1_geometry_fingerprint.pgf}}
\input{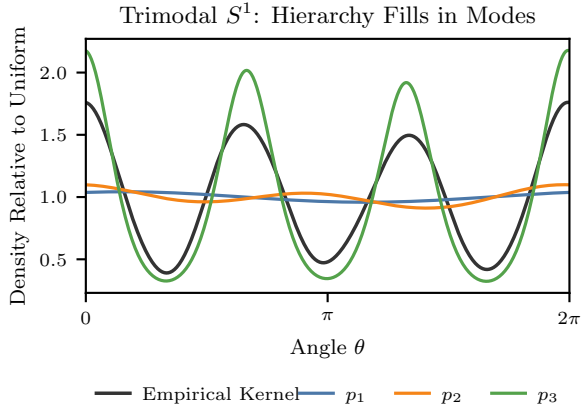}
\caption{Trimodal circle example. The empirical kernel density has three modes. The fits $p_1$ and $p_2$ miss this structure, while $p_3$ captures it. This pattern matches the dominant $I_3$ gap in Table~\ref{tab:s1}.}
\label{fig:s1_hierarchy}
\end{figure}

\paragraph{Trimodal circle hierarchy.}
The first diagnostic is a separate trimodal $S^1$ illustration. We draw $n=500$ angles from the equally weighted mixture
\[
  \frac{1}{3}\sum_{j=0}^{2} \operatorname{vM}\{2\pi j/3,5.5\},
\]
where $\operatorname{vM}(\mu,\kappa)$ is the circular von Mises distribution with mean angle $\mu$ and concentration $\kappa$. All sample weights are uniform. The fitted Fourier maximum-entropy hierarchy uses
\[
  \{\cos(k\theta),\sin(k\theta):1\le k\le L\},
  \qquad L=1,2,3,
\]
with numerical quadrature on 4096 angles. For visualization, the empirical curve is a von Mises kernel density estimate with concentration $14$. Figure~\ref{fig:s1_hierarchy} shows that the first two projections cannot represent the threefold symmetry. The $L=3$ projection recovers the modes. The main-text residuals use the raw $\cos/\sin$ normalization above. An orthonormal Fourier basis multiplies all three by $\sqrt2$ and leaves their comparison unchanged.

\begin{figure*}[ht]
\centering
\input{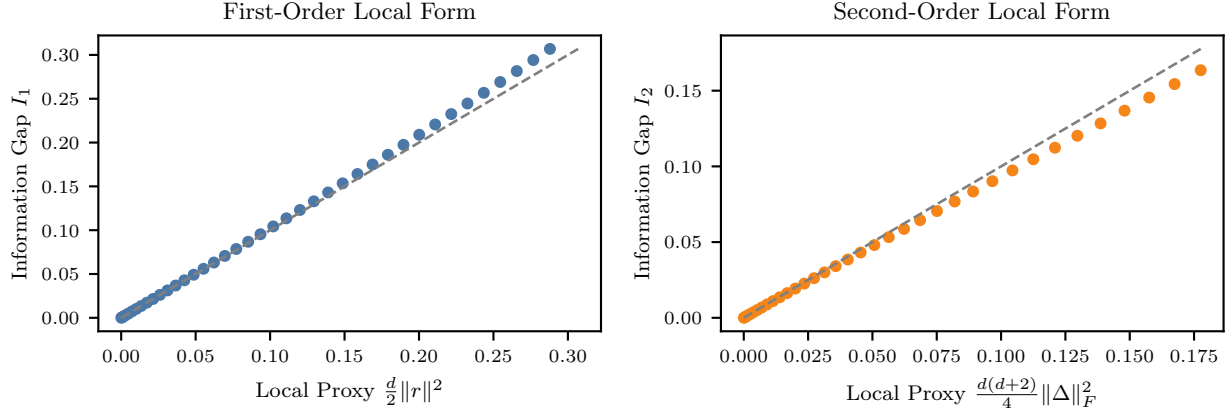}
\caption{Numerical validation of the local interpretations on $S^2$. Left, the first gap agrees with $\frac d2\|r\|^2$ near uniformity. Right, the second gap agrees with $\frac{d(d+2)}4\|Q-I_d/d\|_F^2$ near uniformity. Dashed lines show identity.}
\label{fig:local_forms}
\end{figure*}

\paragraph{Population local-form validation on $S^2$.}
The second diagnostic checks both local interpretations in Section~\ref{sec:sphere}. This calculation is numerical and population-level, not a finite-sample experiment. The first-order panel uses the exact vMF family on $S^2$. Concentration values satisfy $\kappa\in[0.02,1.5]$. For each $\kappa$, the exact entropy deficit is
\[
  I_1(\kappa)=\kappa A_3(\kappa)-\log\{\sinh(\kappa)/\kappa\}
\]
with respect to normalized uniform surface measure. The reference approximation is $(d/2)\|r\|^2$ for $d=3$ and $\|r\|=A_3(\kappa)$. The second-order panel uses an antipodally symmetric Bingham path $p_t(x)\propto\exp\{t x^\top B_0x\}$. Here, $B_0=\operatorname{diag}(1,-1/2,-1/2)$ and $t\in[0.02,1.2]$. Deterministic Fibonacci quadrature approximates the normalizer, second moment $Q_t$, and entropy deficit. The rule has 80,000 points on $S^2$. We compare the second-order deficit with $(d(d+2)/4)\|Q_t-I_d/d\|_F^2$. Figure~\ref{fig:local_forms} confirms both local expansions near uniformity.

\subsection{Additional real-data table}
\label{app:digits_table}

The application uses the handwritten digits dataset from \textsf{scikit-learn} \citep{pedregosa_2011_ScikitlearnMachineLearning}. We standardize the 64-dimensional images, project them onto three principal components, and normalize them to $S^2$. For standardized images $z_i$ and query $z_q$, define
\[
 w_i=\frac{\exp\{6\,z_i^\top z_q/(\|z_i\|\|z_q\|)\}}
 {\sum_j\exp\{6\,z_j^\top z_q/(\|z_j\|\|z_q\|)\}}.
\]
Thus, 6 is an inverse temperature or concentration. We use the first observation from each queried class and retain the query in the reference set. Both levels use a 30,000-point Fibonacci rule. This workflow uses a spherical projection rather than full embedding-scale harmonic fitting.

\begin{table}[ht]
\centering
\caption{Query-weighted digit embeddings after projection to $S^2$. Class mass lists the three largest label masses induced by the query weights.}
\label{tab:digits}
\scriptsize
\resizebox{\columnwidth}{!}{%
\begin{tabular}{@{}lrrrrl@{}}
\toprule
Query & Kish ESS & $I_1$ & $I_2$ & $D_2$ & class mass\\
\midrule
0 & 162.6 & 1.695 & 0.276 & 1.970 & 0:0.85, 9:0.05, 4:0.03\\
1 & 138.2 & 1.166 & 0.153 & 1.319 & 1:0.72, 8:0.06, 4:0.06\\
8 & 85.5 & 0.248 & 0.032 & 0.280 & 8:0.46, 3:0.12, 9:0.11\\
\bottomrule
\end{tabular}
}
\end{table}

At the first two fitted levels, Table~\ref{tab:digits} shows that query ``0'' has a concentrated neighborhood with residual second-order anisotropy. Query ``8'' has less cumulative information. Its lower weight ESS describes weight concentration rather than geometric dispersion. We make no adequacy claim without a residual diagnostic.

% =============================================================

\section{DIAGNOSTICS, REPORTING, AND REGULARIZATION}
\label{app:diagnostics_reporting}
\label{app:adequacy_regularization}

\subsection{Model adequacy and residual diagnostics}

Entropy gaps describe information explained by the selected hierarchy. They do not establish adequacy at level $L$. A low-order model can have high fitted entropy for two reasons. The distribution may be diffuse, or the features may miss relevant structure. A residual diagnostic should therefore accompany the gaps.

One option is an MMD residual. For a characteristic kernel $k$ on $M$, define
\begin{equation*}
  R_L^2(\widehat P_w)
  =\mathrm{MMD}_k^2(\widehat P_w,\widehat p_L\nu).  
\end{equation*}
On the sphere, use the real orthonormal harmonics above with a finite cutoff $K>L$. Write
\[
 \widehat c_{\ell r}=\sum_iw_iY_{\ell r}(x_i),\qquad
 c_{\ell r,L}=\int Y_{\ell r}\widehat p_L\,d\sigma.
\]
Then
\begin{equation}
  R_{L,K}^2(\widehat P_w)
  =\sum_{\ell=L+1}^{K} a_\ell\sum_{r=1}^{h_\ell}
  \left|\widehat c_{\ell r}-c_{\ell r,L}\right|^2.
  \label{eq:harmonic_residual}
\end{equation}
For finite $K$, any $a_\ell\ge0$ may be used, including $a_\ell=1$ or growing Sobolev weights. An infinite-cutoff empirical residual requires summable kernel weights. For example,
\[
 a_\ell=\{1+\ell(\ell+d-2)\}^{-s},\qquad s>(d-1)/2,
\]
ensures $\sum_\ell a_\ell h_\ell<\infty$. Unit or increasing weights therefore cannot be used with the infinite series. A large residual indicates that $U_L$ is not a complete uncertainty summary. Its practical threshold requires separate calibration.

\subsection{Reporting details}
\label{app:reporting}

We recommend reporting gaps $\widehat I_\ell$, cumulative uncertainty $\widehat U_L$, residuals $R_{L,K}$, and Kish weight ESS. Label the ESS as a weight-concentration summary rather than a universal sampling ESS. Reports should identify the sampling mechanism, covariance estimator, feature spaces, projections or sketches, optimization tolerances, quadrature size, seeds, and any shrinkage. Under the correctly specified i.i.d. uniform-weight null, $a_n^2\widehat I_L\Rightarrow\chi^2_{q_L-q_{L-1}}/2$. Its reference mean is $(q_L-q_{L-1})/2$. Thus, raw gaps of different dimensions should not be compared inferentially.

A parametric reduced-model bootstrap is valid only under correct i.i.d. model specification. Under a general moment null, one may instead use a residualized multiplier bootstrap that reproduces the sandwich covariance. Importance-sampling bootstraps must resample proposal draws and recompute weights. Survey procedures must reproduce the sampling design. A prespecified analysis may apply Holm correction to valid level-wise $p$-values from \eqref{eq:weighted_chisq_sim}. The residual threshold and resulting order rule remain heuristic. Inference must account for data-driven projections, feature spaces, or sketches. Sample splitting is one option. Repeated sketches also require multiplicity control. Our experiments report descriptive gaps. Optimal order selection remains separate.

\subsection{Boundary and regularization}

If $\widehat m_L\in\partial\M_L$, no finite natural parameter may solve $\nabla\psi_L(\lambda)=\widehat m_L$. Only a law supported on the preimage of a face of the moment body can match such a moment. The exact GID functionals are then undefined.

For descriptive analysis, a coordinate-free remedy mixes the empirical law with the reference law:
\[
\begin{aligned}
 \widehat P_{w,\epsilon}&=(1-\epsilon)\widehat P_w+\epsilon\nu,
 &0&<\epsilon<1,\\
 \widehat m_{L,\epsilon}&=(1-\epsilon)\widehat m_L.
\end{aligned}
\]
Minimality and full support of $\nu$ imply $0=m_L(\nu)\in\ri(\M_L)$. A strict mixture of any feasible moment with this point is interior. Using the same $\epsilon$ at every level produces finite projections that match the shrunken moments exactly. It preserves nesting, the KL-gap identity, nonnegative gaps, basis invariance, and isometry invariance. The mixture changes the target measure. Therefore, report $\epsilon$ and a sensitivity analysis. Inferential use must incorporate the shrinkage rate into the asymptotics.

In contrast, ridge maximizes $\lambda^\top\widehat m_L-\psi_L(\lambda)-\alpha\|\lambda\|^2/2$ and satisfies $\nabla\psi_L(\lambda)+\alpha\lambda=\widehat m_L$. It does not match moments exactly. Entropy differences from separately penalized fits need not be nonnegative or equal KL gaps. Euclidean ridge also depends on feature coordinates and can break isometry invariance. Ridge may diagnose or stabilize optimization. Its outputs are not exact GID and are not covered by the theorems.

% =============================================================

\section{PROOFS}\label{app:proofs}

\subsection{Existence of the maximum-entropy projection}

Consider the standard finite-dimensional exponential-family argument. Let $\phi:M\to\R^m$ be continuous and define
\[
  \M=\conv\{\phi(x):x\in M\}.
\]
Compactness of $M$ and continuity of $\phi$ make $\M$ compact. Define
\[
  \psi(\lambda)=\log\int_M \exp\{\lambda^\top\phi(x)\}\,d\nu(x).
\]
The gradient and Hessian are
\begin{align*}
  \nabla\psi(\lambda)&=\E_\lambda[\phi(X)],\\
  \nabla^2\psi(\lambda)&=\mathrm{Var}_\lambda\{\phi(X)\}.
\end{align*}
Remove directions $a$ for which $a^\top\phi(x)$ is constant $\nu$-almost surely. Riemannian volume has full support, and $\phi$ is continuous. Thus, this step removes every affine dependence on the support. The remaining Hessian is positive definite, so $\psi$ is strictly convex. Compactness makes $\psi$ finite on the full natural parameter space $\R^m$. The family is regular, and the steepness boundary condition is vacuous \citep{brown_1986_FundamentalsStatisticalExponential}. Standard theory maps $\R^m$ diffeomorphically onto $\ri(\M)$ through $\nabla\psi$. Each $m\in\ri(\M)$ therefore has a unique natural parameter $\lambda(m)$.

The density
\[
  p_m(x)=\exp\{\lambda(m)^\top\phi(x)-\psi(\lambda(m))\}
\]
is feasible. For any feasible density $q$,
\begin{align*}
  \KL(q\nu\,\|\,p_m\nu)
  &=\int q\log q\,d\nu-\int q\log p_m\,d\nu\\
  &=\int q\log q\,d\nu-
  \lambda(m)^\top m+\psi(\lambda(m)).
\end{align*}
The final two terms are fixed over the constraint set. Thus, minimizing $\int q\log q\,d\nu$ is equivalent to minimizing $\KL(q\nu\,\|\,p_m\nu)$. The KL divergence is uniquely minimized by $q=p_m$. Hence, $p_m$ uniquely maximizes entropy.

\subsection{Proof of Theorem \ref{thm:kl_gap}}

Nested features make the level-$L$ constraint set a subset of the level-$(L-1)$ constraint set. Restricting the feasible set cannot increase maximum entropy. Since $D_L=-h(p_L^P)$, we obtain
\[
  D_0(P)\le D_1(P)\le\cdots\le D_L(P).
\]

To prove the KL-gap identity, write the level-$(L-1)$ density as
\[
  \log p_{L-1}^P(x)=\lambda_{L-1}^\top\phi_{L-1}(x)-\psi_{L-1}(\lambda_{L-1}).
\]
The level-$L$ constraints include those at level $L-1$. Therefore, $p_L^P$ and $p_{L-1}^P$ have identical $\phi_{L-1}$ moments:
\[
  \int \phi_{L-1}p_L^P\,d\nu
  =\int \phi_{L-1}p_{L-1}^P\,d\nu
  =m_{L-1}(P).
\]
Therefore
\begin{align*}
  \int p_L^P\log p_{L-1}^P\,d\nu
  &=\lambda_{L-1}^\top m_{L-1}(P)-\psi_{L-1}(\lambda_{L-1})\\
  &=\int p_{L-1}^P\log p_{L-1}^P\,d\nu.
\end{align*}
It follows that
\begin{align*}
 \KL(p_L^P\nu\,\|\,p_{L-1}^P\nu)
 &=D_L(P)-\int p_L^P\log p_{L-1}^P\,d\nu\\
 &=D_L(P)-D_{L-1}(P).
\end{align*}
This proves the identity. Summing over $\ell$ gives \eqref{eq:sum_gaps}.

\subsection{Proof of Theorem \ref{thm:basis_invariance}}

Let $\widetilde\phi=T\phi$, with $T$ invertible. The constraint
\[
  \int \widetilde\phi p\,d\nu=\int \widetilde\phi\,dP
\]
becomes
\[
  T\int \phi p\,d\nu=T\int\phi\,dP.
\]
This is equivalent to the original constraint. Hence, the primal feasible set is unchanged. The projection $p_L^P$ and all entropy-based quantities are also unchanged.

\subsection{Proof of Theorem \ref{thm:isometry_invariance}}

Let $g$ be an isometry. It preserves normalized Riemannian volume. Thus, $p\mapsto p\circ g^{-1}$ is an entropy-preserving bijection between densities. Suppose $\V_L$ is invariant under the group action. Moment matching in $\V_L$ under $P$ corresponds to moment matching under $g_\#P$. Hence, the projected measure is the corresponding pushforward:
\[
  p_L^{g_\#P}\nu=g_\#(p_L^P\nu).
\]
KL divergence to $\nu$ is invariant under measure-preserving transformations. Therefore, $D_L(g_\#P)=D_L(P)$. The results for $U_L$ and $I_L$ follow.

\subsection{Proof of Theorem \ref{thm:consistency}}

Assumption \ref{assump:interior} places $m_L(P)$ in $\ri(\M_L)$ and makes the mean map continuous near $m_L(P)$. Assumption \ref{assump:moment_conv} gives $\widehat m_L\to m_L(P)$ in probability. Therefore, $\widehat m_L\in\ri(\M_L)$ with probability tending to one. Consider all plug-in quantities on this event. The continuous mapping theorem gives
\[
  \widehat\lambda_L=\lambda_L(\widehat m_L)\xrightarrow{p}\lambda_L(m_L(P)).
\]
The entropy deficit can be written as the Legendre transform
\[
  D_L(m)=\lambda_L(m)^\top m-\psi_L(\lambda_L(m)).
\]
This transform is continuous on the moment body's relative interior. Therefore, $\widehat D_L\to D_L(P)$ in probability. Consistency of $\widehat I_L=\widehat D_L-\widehat D_{L-1}$ follows.

\subsection{Proof of Theorem \ref{thm:clt}}

The entropy deficit $D_L(m)$ is the convex conjugate of $\psi_L$ on the feasible moment body. It is differentiable on the relative interior, with
\[
  \nabla_mD_L(m)=\lambda_L(m).
\]
Applying the delta method to the moment CLT gives
\[
  a_n(\widehat D_L-D_L)
  \Rightarrow N(0,\lambda_L^\top\Sigma_L\lambda_L).
\]
For the increment, view $D_{L-1}$ as a function of level-$L$ moments through $\Pi_{L-1}m_L$. Its gradient in level-$L$ coordinates is $\Pi_{L-1}^\top\lambda_{L-1}$. Thus, the gradient of $I_L=D_L-D_{L-1}$ is
\[
  \gamma_L=\lambda_L-\Pi_{L-1}^\top\lambda_{L-1}.
\]
A second application of the delta method proves \eqref{eq:I_clt}.

\subsection{Proof of Proposition \ref{prop:I1_local}}

Under the uniform law on $S^{d-1}$, $\E[X]=0$ and $\mathrm{Var}(X)=I_d/d$. The level-1 exponential tilt is $p_1(x)\propto\exp(\eta^\top x)$. For small $\eta$,
\[
  r(P)=\E_\eta[X]=\frac{\eta}{d}+o(\norm{\eta}).
\]
The entropy deficit has the local Fisher expansion
\[
  D_1=\frac12\eta^\top \mathrm{Var}_0(X)\eta+o(\norm{\eta}^2)
  =\frac{\norm{\eta}^2}{2d}+o(\norm{\eta}^2).
\]
Substituting $\eta=d r(P)+o(\norm{r(P)})$ gives
\[
  I_1(P)=D_1(P)=\frac d2\norm{r(P)}^2+o\{\norm{r(P)}^2\}.
\]

\subsection{Proof of Proposition \ref{prop:I2_local}}

For $X$ uniform on $S^{d-1}$,
\[
  \E[X_iX_jX_kX_l]
  =\frac{\delta_{ij}\delta_{kl}+\delta_{ik}\delta_{jl}+\delta_{il}\delta_{jk}}{d(d+2)}.
\]
For a traceless symmetric matrix $B$, this gives
\begin{align*}
  \E\{(X^\top BX)XX^\top\}&=\frac{2B}{d(d+2)},\\
  \E\{(X^\top BX)^2\}&=\frac{2\tr(B^2)}{d(d+2)}.
\end{align*}
For fixed $B$, the log-partition is even and strictly convex in $\eta$. Its $\eta$-gradient therefore vanishes only at $\eta=0$. The projection matches $\E_PX=0$, so its linear parameter is zero. Near uniformity, its quadratic parameter and $\Delta=Q(P)-I_d/d$ satisfy
\[
  \Delta=\frac{2B}{d(d+2)}+o(\norm{B}_F).
\]
The entropy deficit of the corresponding quadratic exponential tilt is
\[
\begin{aligned}
 D_2&=\tfrac12\E\{(X^\top BX)^2\}+o(\norm{B}_F^2)\\
 &=\frac{\tr(B^2)}{d(d+2)}+o(\norm{B}_F^2).
\end{aligned}
\]
Since the first moment is zero, $D_1=0$ and $I_2=D_2$. Substituting
$B=\frac{d(d+2)}{2}\Delta+o(\norm{\Delta}_F)$ yields
\[
  I_2(P)=\frac{d(d+2)}{4}\norm{\Delta}_F^2+o(\norm{\Delta}_F^2),
\]
as claimed.

\subsection{Proof of Theorem \ref{thm:null}}

Write $m=(m_u,m_v)$ for the level-$L$ moment vector. Under $I_L(P)=0$, the full and reduced projections coincide. Their common natural parameter is $\theta_0=(\alpha_0,0)$. Let
\[
  \mathcal J_0=\nabla^2\psi_L(\theta_0)=
  \begin{pmatrix}J_{uu}&J_{uv}\\J_{vu}&J_{vv}\end{pmatrix}.
\]
Minimality implies $\mathcal J_0\succ0$ and $J_{uu}\succ0$. The Schur complement $\mathcal S=J_{vv}-J_{vu}J_{uu}^{-1}J_{uv}$ is also positive definite. At the null, the Hessian of $D_L(m)$ is $\mathcal J_0^{-1}$. The reduced deficit $D_{L-1}(m_u)$ has Hessian $J_{uu}^{-1}$ in the $u$ coordinates. Thus, the Hessian of $I_L(m_u,m_v)=D_L(m_u,m_v)-D_{L-1}(m_u)$ is
\[
  \mathcal J_0^{-1}-
  \begin{pmatrix}J_{uu}^{-1}&0\\0&0\end{pmatrix}.
\]
Using the block inverse formula, this difference equals
\[
  \mathcal{N}_0^\top \mathcal S^{-1}\mathcal{N}_0,
  \qquad
  \mathcal{N}_0=\begin{pmatrix}-J_{vu}J_{uu}^{-1}&I_q\end{pmatrix}.
\]
Since $I_L(P)=0$ and $\nabla I_L=0$ at the null, the second-order Taylor expansion gives
\[
  \widehat I_L=\frac12(\widehat m_L-m_L)^\top \mathcal{N}_0^\top \mathcal S^{-1}\mathcal{N}_0(\widehat m_L-m_L)+o_p(a_n^{-2}).
\]
Multiplying by $a_n^2$ and applying the moment CLT proves \eqref{eq:null_quad}. Now suppose i.i.d. samples follow the correctly specified reduced model with uniform weights. Then $\Sigma_L=\mathcal J_0$ and $a_n=\sqrt n$. Also, $\mathcal{N}_0\Sigma_L\mathcal{N}_0^\top=\mathcal S$. Consequently, $2a_n^2\widehat I_L\Rightarrow\chi_q^2$.

\subsection{CLT for independent samples with deterministic weights}

Suppose $X_i\overset{\mathrm{i.i.d.}}\sim P$, the weights are deterministic, and $\sum_iw_i=1$. Let $Z_i=\phi_L(X_i)-m_L(P)$. Then
\[
  \widehat m_L-m_L(P)=\sum_iw_iZ_i.
\]
If $\E\norm{Z_i}^2<\infty$ and the Lindeberg condition
\[
  \frac{\max_iw_i}{(\sum_jw_j^2)^{1/2}}\to0
\]
holds, then
\[
  \frac{\widehat m_L-m_L(P)}{(\sum_iw_i^2)^{1/2}}
  \Rightarrow N(0,\mathrm{Var}_P\{\phi_L(X)\}).
\]
Compactness of $M$ and continuity of $\phi_L$ imply the moment condition. This identifies $a_n=(\sum_iw_i^2)^{-1/2}$ and the Kish weight ESS only for this i.i.d., deterministic-weight setting.

\subsection{Self-normalized importance sampling and other designs}

For the main-text influence covariance, suppose $X_i\overset{\mathrm{i.i.d.}}\sim Q$, $P\ll Q$, and the exact density ratio $r=dP/dQ$ is available. With $m=m_L(P)$, the self-normalized estimator has influence function
\[
 \frac{r(X)}{\E_Qr(X)}\{\phi_L(X)-m\}.
\]
The usual $\sqrt n$ CLT and its empirical covariance require
\[
 \E_Q\!\left[r(X)^2\|\phi_L(X)-m\|^2\right]<\infty
\]
and consistency of the estimated null mean. Adaptive or estimated proposals and MCMC particles require covariance estimates that reflect their dependence. Attention weights, reliability weights, and complex surveys require covariance estimates reflecting their dependence or sampling design. Inverse squared weights alone do not determine inferential precision.

\end{document}